\documentclass[aps,pre,twocolumn,amsmath,showkeys,nofootinbib,amssymb,amsfonts]{revtex4}
\usepackage{dcolumn}

\usepackage[pdftex]{graphicx}
\usepackage{bm}
\begin{document}
\author{Michele Caponigro}
\affiliation{ISHTAR  Indeterminism in Sciences and Historico-philosophical\\ Transdisciplinar Advanced Research Center,
University of Bergamo} \email{michele.caponigro@unibg.it}
\author{Enrico Giannetto}
\affiliation{Department of Human Sciences, University of Bergamo} \email{enrico.giannetto@unibg.it}
\title{Epistemic vs Ontic Classification of quantum entangled states?}
\keywords{Quantum entanglement, subsystems (partitions and factorizables states), epistemic vs ontic elements}
\begin{abstract}
In this brief paper, starting from recent works, we analyze from a conceptual point of view this basic question: can the
nature of quantum entangled states be interpreted ontologically or epistemologically? According to some works, the
degrees of freedom (and the tool of quantum partitions) of quantum systems permit us to establish a possible
classification between factorizable and entangled states. We suggest, that the "choice" of degree of freedom (or quantum
partitions), even if mathematically justified introduces an epistemic element, not only in the systems but also in their
classification. We retain, instead, that there are not two classes of quantum states, entangled and factorizable, but
only a single class of states: the entangled states. In fact, the factorizable states become entangled for a different
choice of their degrees of freedom (i.e. they are entangled with respect to other observables). In the same way, there
are no partitions of quantum systems which have an ontologically superior status with respect to any other. For all these
reasons, both mathematical tools utilize(i.e quantum partitions or degrees of freedom) are responsible for creating an
improper classification of quantum systems. Finally, we argue that we cannot speak about a classification of quantum
systems: all quantum states exhibit a uniquely objective nature, they are all entangled states
\end{abstract}
\maketitle
\section{Systems and Partitions}
In spite of continuous progress, the current state of entanglement theory is still marked by a number of outstanding
unresolved problems. These problems range from the complete classification of mixed-state bipartite entanglement to
entanglement in systems with continuous degrees of freedom, and the classification and quantification of multipartite
entanglement for arbitrary quantum states.\\
In this paper, starting form two important works,\\
1) Torre (2010) and 2) Zanardi (2001), we will analyze the possible relationship among these elements:
\begin{enumerate}
\item the degrees of freedom of the quantum system
\item the partitions of the quantum system
\item the epistemic elements introduced from the procedures (1) and (2)
\end{enumerate}
As we know,  the relationship between quantum systems (QS) and their possible quantum entangled systems (QES) is not a
trivial question. There are many efforts to understand this dynamics. Zanardi (2001) in his paper argues that the
partitions of a possible system do not have an ontologically superior status with respect to any other: according Zanardi
given a physical system $S$, the way to subdivide it in subsystems is in general by no means unique. We will analyze his
conclusion in the following sections below.\\
According Zanardi the consequences of the \textbf{non uniqueness of the decomposition of a given system $S$ into
subsystems} imply (at the quantum level), a fundamental ambiguity about the very notion of entanglement that accordingly
becomes a relative one. The concept of "relative" for an entangled system, has been developed by Viola and Barnun (2006).
They concentrate their efforts on this fundamental question: how can entanglement be understood in an arbitrary physical
system, subject to arbitrary constraints on the possible operations one may perform for describing, manipulating, and
observing its states? In their papers, the authors proposed that entanglement is an inherently \textbf{relative concept},
whose essential features may be captured in general in terms of the relationships between different observers (i.e.
expectations of quantum observables in different, physically relevant sets). They stressed how the the role of the
\textbf{observer} must be properly acknowledged in determining the distinction between entangled and unentangled states.
\subsection{Quantum Entanglement: brief overview}
From a phenomenological point of view, the phenomenon of entanglement is quite simple. When two or more physical systems
form an interaction, some correlation of a quantum nature is generated between the two of them, which persists even when
the interaction is switched off and the two systems are spatially separated. Quantum entanglement describes a
non-separable state of two or more quantum objects and has certain properties which contradict common physical sense.
While the concept of entanglement between two quantum systems, which was introduced by E. Schr\"{o}dinger (1936) is well
understood, its generation and analysis still represent a substantial challenge. Moreover, the problem of quantification
of entangled states, is a long standing issue debated in quantum information theory. Today the bipartite entanglement
(\textbf{two-level systems, i.e. qubits}) is well understood and has been prepared in many different physical systems.
The mathematical definition of entanglement varies depending on whether we consider only pure states or a general set of
mixed states (see Giannetto 1995: where it is discussed the reason why entanglement generally requires a density matrix
formalism). In the case of pure states, we say that a given a state $|\psi\rangle$ of $n$ parties is \emph{entangled} if
it is not a tensor product of individual states for each one of the parties, that is,
\begin{equation}
|\psi \rangle \ne |v_1\rangle_1 \otimes |v_2 \rangle_2 \otimes
\cdots \otimes |v_n\rangle_n \ .
\end{equation}
For instance, in the case of $2$ qubits $A$ and $B$ (sometimes called "Alice" and "Bob") the quantum state
\begin{equation}
|\psi^+\rangle = \frac{1}{\sqrt{2}} [\left( |0\rangle_A \otimes
|0\rangle_B + |1\rangle_A \otimes |1\rangle_B \right)]
\label{maxen}
\end{equation}
is entangled since $|\psi^+\rangle \ne |v_A\rangle_A \otimes |v_B\rangle_B$. On the contrary, the state
\begin{equation}
|\phi\rangle=\frac{1}{2}[( |0\rangle_A \otimes |0\rangle_B +
|1\rangle_A \otimes |0\rangle_B +|0\rangle_A \otimes |1\rangle_B +
|1\rangle_A \otimes |1\rangle_B )]
\end{equation}
is not entangled, since
\begin{equation}
|\phi\rangle = \left(\frac{1}{\sqrt{2}}\left( |0\rangle_A +
|1\rangle_A \right) \right) \otimes \left(\frac{1}{\sqrt{2}}\left(
|0\rangle_B + |1\rangle_B \right) \right) \ . \label{separab}
\end{equation}
A pure state like the one from Eq.2 is called a \emph{maximally entangled state of two qubits}, or a \emph{Bell pair},
whereas a pure state like the one from Eq.4 is called \emph{separable}. In the general case of mixed states, we say that
a given state $\rho$ of $n$ constituent states is \emph{entangled} if it is not  a probabilistic sum of tensor products
of individual states for each one of the subconstituents, that is,
\begin{equation}
\rho \ne \sum_k p_k  \ \rho^k_1 \otimes \rho^k_2 \otimes \cdots
\otimes \rho^k_n \ ,
\end{equation}
with $\{ p_k \}$ being some probability distribution. Otherwise, the mixed state is called \emph{separable}. The essence
of the above definition of entanglement relies on the fact that entangled states of $n$ constituents cannot be prepared
by acting locally on each one of them, together with classical communication among them. Entanglement is a genuinely
quantum-mechanical feature which does not exist in the classical world. It carries non-local correlations between the
different systems in such a way that they cannot be described classically.
\section{Are quantum states all Entangled?}
As mentioned above, the recent work by Torre (2010) is a fundamental paper which gives us the possibility to speculate
about the nature and the classification of entangled states. The paper demonstrates that a state is factorizable in the
Hilbert space corresponding to some choice of degrees of freedom, and that this same state becomes entangled for a
different choice of degrees of freedom. Therefore, entanglement is not a special case, but is ubiquitous in quantum
systems. According to the authors, one may erroneously think that there are two classes of states for the QS: 1)
factorizable and 2) entangled, which correspond to qualitative difference in the behaviour of the system, close to
classical in one case and with strong quantum correlations in the other. They argue that this is indeed wrong because
factorizable states also exhibit entanglement with respect to \emph{other} observables. In this sense, all states are
entangled; \textbf{entanglement is not an exceptional feature of some states but is ubiquitous in QM.}.\\
To sum up this conceptual analysis by Torre and Zanardi, we think that there is an unclear relationship among these
elements:
\begin{enumerate}
\item factorizable states (Torre)
\item entangled states
\item the (choice) of partitions of quantum system (Zanardi)
\item the role of the observer (in determining the distinction between entangled and unentangled states)
\end{enumerate}
We think that all points (except the second point) introduce epistemic elements in the analysis and in the classification
of the quantum systems. We suggest that the second point is the key to understand the nature of the underlying physical
reality. We argue in the next sections, that the conceptual analysis of Torre and Zanardi differs from what we suggest
concerning the epistemic elements introduced in their papers.
\subsection{Factorizability of a state as Epistemic property?}
\begin{figure}
\includegraphics[width=8cm,height=2cm]{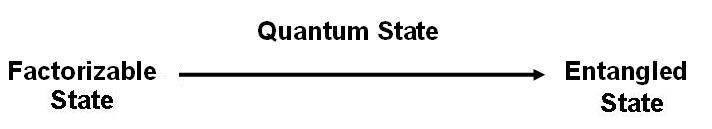}
\caption {Torre's main thesis: factorizable states
become entangled in a different degrees of freedom.}
\end{figure}
An important question is related at the property of factorizability of quantum state. Is the factorizability tool an
objective property? Briefly stored, is factorizability an objective property of the system or is it a feature of (our)
description of system (i.e. an epistemic property)?. With reference to Torre's paper (2010), the authors show that
factorizability and entanglement \textbf{are not preserved} in a change of the degrees of freedom used to describe the
system, they demonstrate in detailed case that the factorizability of a state is a property that is \textbf{not}
invariant under a change of the degrees of freedom that we use in order to describe the system. From mathematical point
of view\cite{Tor}, they consider a quantum system with two subsystems $S=(S_{A},S_{B})$ that may correspond to two
degrees of freedom $A$ and $B$. The state of the system belongs then to the Hilbert space
$\mathcal{H}=\mathcal{H}_{A}\otimes\mathcal{H}_{B}$ and the two degrees of freedom are represented by operators $A\otimes
\mathbb{I} $ and $\mathbb{I}\otimes  B $. Suppose it is given that the system has a factorizable, non entangled, state
$\Psi=\Psi_{A}\otimes\Psi_{B}$ with $\Psi_{A}$ and $\Psi_{B}$ arbitrary states (not necessarily eigenvectors of $A$ and
$B$) in the spaces $\mathcal{H}_{A}$ and $\mathcal{H}_{B}$. Then there exists a transformation of the degrees of freedom
$F=F(A,B)$ and $G=G(A,B)$ that suggests a different factorization, $\mathcal{H}=\mathcal{H}_{F}\otimes\mathcal{H}_{G}$,
where the state is no longer factorizable: $\Psi\neq\Psi_{F}\otimes\Psi_{G}$ with $\Psi_{F}\in\mathcal{H}_{F}$ and
$\Psi_{G}\in\mathcal{H}_{G}$. The state becomes entangled with respect to the new degrees of freedom; the factorizability
of \emph{states} is not invariant under a different factorization of the Hilbert space. To conclude, they have shown that
for any system in a factorizable state, it is possible to find different degrees of freedom that suggest a different
factorization of the Hilbert space where the same state becomes entangled; for this reason they argued that every state,
even for those factorizable, it is possible to find pairs of observables that will violate Bell's inequalities. The figure above (n.1 pag.2) summarize Torre's thesis.\\
The authors analyze also the inverse problem: the fact that the appearance of entanglement depends on the choice of
degrees of freedom can find an interesting application in the "disentanglement" of a state; one can, sometimes, transform
an entangled state into a factorizable one by a judicious choice of the degrees of freedom.\\
To conclude, we think that the epistemic element is inherent in the possibility to "choose" the degrees of freedom of the
quantum system: this possibility affects the classification of quantum states in entangled or factorizables. In fact, it
is simple to ask these epistemological questions: a)what are the degrees of freedom for a quantum system? b) Is it a
complete set that describe all quantum properties? Can be a particle entangled in one context be factorizables in another
context?
\subsection{The partitions of quantum system as Epistemic property?}
As we have seen, given a quantum system, the way to subdivide (to partition) it in subsystems in \textbf{not unique}. We
call this first phase "epistemic", as in fact we are able to decide how to partition the quantum system. The conclusion
of this operation is most important of its premise: in fact if we find (in the subsystems) an entangled state, this state
has an ontological nature \textbf{but only if referred} to that kind of particular partition carried. We have, in other
words, an objective entangled state for an epistemic partitions! For these reasons, the notions of an entangled state
becomes a relative concept and the relativity of this concept is linked, to us, at the choice of partitions or degrees of
freedom. At the same time, the property of the entangled state is objective. The figure above (n.2 pag.3) represent our
view of Zanardi's problem.
\begin{figure}[!htbp]
\includegraphics[height=5.5cm,width=7cm]{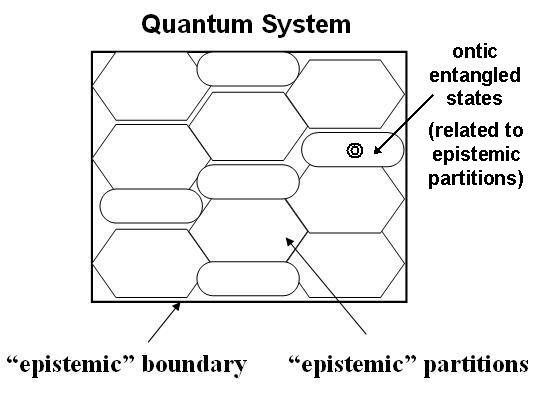}
  \caption {Epistemic Partitions and Ontic entanglement}
\end{figure}
\section{Some Considerations and conclusions}
We have seen that quantum systems admit a variety of tensor product structures depending on the complete system of
commuting observables chosen for the analysis; as a consequence we have different notions of entanglement associated with
these different tensor product.
\begin{figure}[h!]
\includegraphics[height=3.5cm,width=8cm]{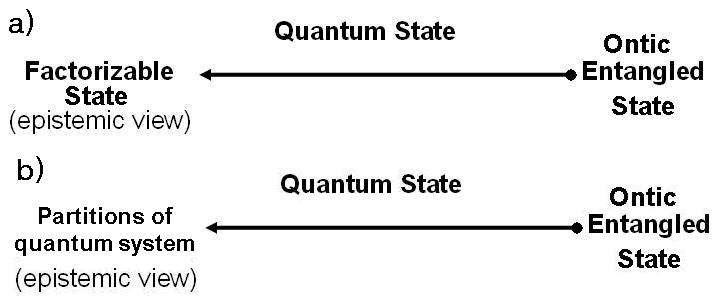}
  \caption {Our position}
\end{figure}
We notice that, in the determination of whether a state is factorizable or entangled, the factorization of the Hilbert
space is crucial and this factorization depends on the choice of the observables corresponding to the degrees of freedom.
In the same way, as Zanardi stressed, given a quantum system, the way to subdivide it (via partitions) in subsystems it
is \textbf{not unique}; the partitions of a possible system have not an ontologically superior status with respect to any
other. Based on these points, we argue, that the criteria of partitions and factorizability (or partitions) contain an
a-priori epistemic element, the figure n.3 (pag.3)summarize our position. In conclusion, we suggest that all quantum
system exhibit an objective nature that is entangled, at basic level the underlying physical reality is entangled. A
quantum state could be non-entangled if and only if it would be factorizable for every possible partition or choice of
degrees of freedom, but this can never occur. The epistemic level emerges with the "observer" (partitions or degree of
freedom), the physicists and philosophers should consider these arguments in their debates.

\end{document}